\documentclass{ws-ijmpd}
\usepackage{epsfig}

\newcommand{\uk}{u_k}

\begin{document}
\newcommand{\bbi}[1]{\mbox{\boldmath $#1$}}
\newcommand{\bay}[2]{\left(\begin{array}{c}\mbox{$#1$} \\ \mbox{$#2$} \end{array}\right)} 
\markboth{Lung-Yih Chiang}
{Non-linearity of large-scale structure formation in the Universe}

\catchline{}{}{}{}{}

\title{Non-linearity of large-scale structure formation in the Universe}

\author{Lung-Yih Chiang}

\address{Niels Bohr Institute, \\
 Blegdamsvej 17, DK-2100 Copenhagen, Denmark\\
chiang@nbi.dk}

\maketitle

\begin{abstract}
In the standard picture of cosmological structure formation, the Universe we see today is evolved under the gravitational instability from tiny random fluctuations. In this talk I discuss the onset of non-linearity in the large scale structure formation of the Universe when the linear perturbation theory break downs. Using 1D Zel'dovich Approximation which provides an exact solution for density evolution, I illustrate two effects: mode spawning and mode merging and their connection to mode coupling. Those mode couplings (quadratic, cubic \ldots etc.) from gravitational clustering are in fact what the polyspectra (bispectrum, trispectrum \ldots etc.) are meant to measure.  
\end{abstract}

\section{Introduction}
One of the main issues in cosmology is quantitative characterization of the large-scale structure of the Universe. Recent large sky surveys such as SDSS (see {\tt http://www.sdss.org} and the publications therein) and 2dFGRS (see {\tt http://www.mso.anu.edu.au/2dFGRS/} and the publications therein) have displayed rich structures of our Universe. In the framework of the inflation paradigm \cite{star,guth,guthpi,alb,linde}, the structure we see today is blown up from tiny inhomogeneities originated from quantum fluctuations. This picture is confirmed by cosmic microwave background (CMB) measurement of NASA WMAP mission \cite{wmap,wmapresults,wmapcl,wmapng,wmappar}. The homogeneity and isotropy with fluctuations only 1 part in $10^5$ of CMB at the epoch of decoupling provides the strong evidence that our Universe was very smooth Such initial density fluctuations in the simplest inflationary universe constitute a Gaussian random field. 

\section{Gaussian random fields and Fourier phases}
Gaussian random Fields \cite{adler,bbks} are useful because of its analytical
simplicity. One particularly interesting property of Gaussian random
fields is that the real and imaginary part of the Fourier modes are
both Gaussian distributed and mutually independent, or in other
words, the Fourier phases are randomly distributed between 0 and $2
\pi$. The statistical properties are then completely specified by its
second-order statistics: its two-point correlation function $\xi(r)$
(with zero connected $n$-point correlation functions for $n>2$), or
alternatively, its power spectrum $P(k)$. 

In the framework of gravitational instability, a perturbative method
can be adopted at the early stage of density clustering (see
e.g. Peebles 1980, Sahni and Coles 1995, Bernardeau et al. 2002). The linear perturbation
theory is applicable when the density fluctuations are small i.e. its
variance of density contrast $\langle \delta^2 \rangle <1$. The
statistical distribution of the density field such as originally
random phases remains invariant in the linear regime, except its
variance increasing with time. 

The departure of the density field from the linear regime gives
rise to phase coupling. Second-order statistics, such as power
spectrum and two-point correlation function, throw
away the fine details of the delicate pattern of cosmic
structure. These details lie in the distribution of Fourier phases to
which second-order statistics are blind. The evident shortcomings of
$P(k)$ can be partly ameliorated by defining higher-order quantities
such as the bispectrum \cite{peebles}.
or correlations of $\delta({\bbi k})^2$ \cite{stir}. Higher-order
correlations and polyspectra find this information term by term so
that an infinite hierarchy is required for a complete statistical
characterization of the fluctuation field. 

\section{Fourier description}
The mathematical description of an inhomogeneous Universe revolves
around the dimensionless density contrast, $\delta({\bbi x})$,
which is obtained from the spatially-varying matter density
$\rho({\bbi x})$ via
\begin{equation} \delta ({\bbi x}) =
\frac{\rho({\bbi x})-\rho_0}{\rho_0},
\end{equation}
where ${\bbi x}$ is the comoving coordinate, $\rho_0$ is the global
mean density \cite{peebles}. The two-point covariance function, which
measures the excess probability over Poisson distribution between a
pair with separation ${\bbi r}$, is defined as
\begin{equation}
\xi({\bbi r})=\langle \delta({\bbi x}) \delta({\bbi x} +{\bbi r}) \rangle,
\end{equation}
where the mean is taken over all points ${\bbi x}$. It is also useful to
expand the density contrast in Fourier series, in which $\delta$ is
treated as a superposition of plane waves: 
\begin{equation}
\delta ({\bbi x}) = \sum \delta({\bbi k}) \exp(\imath {\bbi k}\cdot
{\bbi x}). \label{eq:fourier}
\end{equation}
The Fourier transform $\delta({\bbi k})$ is complex and
therefore possesses both amplitude $|\delta ({\bbi k})|$
and phase $\phi_{\bbi k}$ where
\begin{equation}
\delta({\bbi k})=|\delta ({\bbi k})| \exp(\imath \phi_{\bbi k}).
\label{eq:fourierex}
\end{equation}
The power spectrum is defined as
\begin{equation}
\langle \delta({\bbi k}_1) \delta({\bbi k}_2) \rangle = (2 \pi)^3 P(k)
\delta^D({\bbi k}_1+{\bbi k}_2), \label{eq:powersp}
\end{equation}
which is the Fourier transform of the two-point covariance function
via Wiener-Khintchin theorem. We can analogously define the bispectrum
as the third-order moment in Fourier space:
\begin{equation}
\langle \delta({\bbi k}_1) \delta({\bbi k}_2) \delta({\bbi k}_3) \rangle =
(2 \pi)^3 B({\bbi k}_1,{\bbi k}_2,{\bbi k}_3) \delta^D({\bbi k}_1+{\bbi
k}_2+{\bbi k}_3 ), \label{eq:bisp}
\end{equation}
which is the Fourier transform of the connected three-point covariance
function.

\section{The linear theory of density perturbations}
Gravitational instabilities is believed to be the driving force of
large-scale structure formation. When the density perturbation is small, evolution of the density contrast can be
obtained analytically through the {\it linear perturbation theory} from 3
coupled partial differential equations. They are the linearized continuity
equation,
\begin{equation}
{\partial\delta\over \partial t} = - {1\over a}{\bbi \nabla_x}\cdot
{\bbi v}, \label{eq:lCont}
\end{equation}
the linearized Euler equation
\begin{equation}
{\partial {\bbi v}\over\partial t} + {\dot a\over a}{\bbi v} = -
{1\over \rho a}{\bbi \nabla_x} p -{1\over a}{\bbi \nabla_x}\phi,
\label{eq:lEuler}
\end{equation}
and the linearized Poisson equation
\begin{equation}
{\bbi \nabla_x}^2\phi = 4\pi G a^2\rho_0\delta. \label{eq:lPoisson}
\end{equation}

In these equations, $a$ is the expansion factor, $p$ is the pressure,
${\bbi \nabla_x}$ denotes a derivative with respect to the comoving
coordinates ${\bbi x}$, ${\bbi v}=a \dot{\bbi x}$ is the peculiar
velocity and  $\phi({\bbi x},t)$ is the peculiar gravitational
potential. From Eq.(\ref{eq:lCont})-(\ref{eq:lPoisson}), and if one
ignores pressure forces, it is easy to obtain an equation for the
evolution of $\delta$:
\begin{equation}
\ddot\delta + 2(\frac{\dot a}{a})\delta - 4 \pi G \rho_0 \delta = 0.
\label{eq:2ndorder}
\end{equation}
For a spatially flat universe dominated by pressureless matter,
$\rho_0(t) = 1/6\pi Gt^2$ and Eq.(\ref{eq:2ndorder}) admits two
linearly independent power-law solutions 
\begin{equation}
\delta({\bbi x},t) = b_{\pm}(t)\delta_0({\bbi x}), \label{eq:linearsol}
\end{equation} 
where $\delta_0({\bbi x})$ is the initial density distribution, $b_+(t) \propto
a(t) \propto t^{2/3}$ is the growing mode and $b_{-}(t) \propto
t^{-1}$ is the decaying mode.

As one can see from Eq.(\ref{eq:linearsol}), the growth depends only
on time, so the growth of density distribution in the linear regime
does not alter the phases.

\section{The Zel'dovich Approximation}
In order to examine analytically non-linear effects induced by gravitational
clustering, we use the Zel'dovich approximation\cite{za} as a clustering
scheme, which extrapolates the evolution of density
perturbations into non-linear regime. This extrapolation from the
linear theory follows the perturbation in particle trajectories
rather than in density fields. In the Zel'dovich approximation (ZA), a
particle initially placed at the Lagrangian coordinate ${\bbi q}$ is
perturbed after a time $t$ to an Eulerian coordinate ${\bbi x}$. The
displacement of the particle simply depends on the constant velocity the
particle has when it is kicked off its initial position, and can be
written as $b(t){\bbi u}({\bbi q})$, so that
\begin{equation}
{\bbi r}({\bbi q},t)=a(t){\bbi x}({\bbi q},t)=a(t)[{\bbi q}+b(t){\bbi u}({\bbi q})],
\end{equation}
where ${\bbi r}$ is the resultant physical coordinate, $a(t)$ is the expansion factor and $b(t)$ is the growing
mode $b_{+}(t)$ from the linear perturbation theory. According to this
prescription, each particle moves with a constant velocity
along a ballistic trajectory, which resembles Newtonian inertial
motion. Note that the peculiar velocity
according to the ZA is $a\dot{\bbi x}=a(t) \dot{b}(t) {\bbi u}({\bbi
q})$. For an irrotational flow, ${\bbi u}({\bbi q})$ can be expressed as
a gradient of some velocity potential $-\nabla \Phi_{0}({\bbi q})$. 

We focus particularly on the applications of one-dimensional ZA. 
The ZA in 1D provides an exact solution of
density evolution in that the evolution of planar collapse from the ZA
has the same solution as from the Poisson equation \cite{buchert}
until shell-crossing. The ZA in 1D is simplified to 
\begin{equation}
x(q,t)=q+b(t) u(q) = q-b(t)\frac{d\Phi_0(q)}{dq}. \label{eq:1d}
\end{equation}
The density contrast can be derived from the conservation of mass $\rho dx=\rho_0 dq$  :
\begin{eqnarray}
\delta & = & \frac{\rho}{\rho_0}-1=\left(\frac{\partial x}{\partial
q}\right)^{-1}-1  = \left[1-b(t)\frac{d^2\Phi_0(q)}{dq^2}
\right]^{-1}-1 \label{eq:delta1d} \\
& = & \sum_n b^n(t)\left( \frac{d^2
\Phi_0}{dq^2}\right)^n. \label{eq:expansion}
\end{eqnarray}
The velocity potential $\Phi_0(q)$ can always be disintegrated as
\begin{equation}
\Phi_0(q)=\sum_i A_i \cos(\lambda_i q + \alpha_i).\label{eq:potential}
\end{equation}

\section{Mode spawning and mode merging}
Upon using 1D ZA, we firstly choose as a toy model the velocity potential \footnote{We put
minus sign for both cosine functions in the velocity potential to make the analytical form neat,
assuming both $A_1$ and $A_2$ positive, otherwise the phases would
have a shift by $\pi$.}
\begin{equation}
\Phi_0(q) = - A_1 \cos(\lambda_1 q + \alpha_1)- A_2 \cos(\lambda_2 q +
\alpha_2).
\end{equation}
When $b(t) \ll 1$, the density contrast can be expressed in terms of
only the first order from  Eq.(\ref{eq:expansion})
\begin{eqnarray}
\delta \;\; \simeq \;\;\delta^{(1)}& = &  b(t)\left( \frac{d^2
\Phi_0}{dq^2}\right) \nonumber \\
&= & a_1 \cos(\lambda_1 q + \alpha_1) + a_2 \cos( \lambda_2 q+ \alpha_2) \nonumber \\
& \equiv & a_1 \bay{\lambda_1}{\alpha_1} + a_2
\bay{\lambda_2}{\alpha_2},
\end{eqnarray}
where the round brackets hereafter denote cosine functions \cite{coles} in order to show clearly the harmonic relationship (see below), $a_1= b(t) A_1
\lambda^2_1$ and $a_2= b(t) A_2 \lambda^2_2$. After Fourier transform
in Lagrangian coordinate we have only two modes with wavenumbers
$\lambda_1$ and $\lambda_2$, and phases $\alpha_1$ and $\alpha_2$,
respectively. So these are the only Fourier modes the moment
the clustering process begins (which we call the ``parent modes''). It is clear that the first order does not
display phase coupling. When the second-order term becomes comparable,   
\begin{eqnarray}
\delta & \simeq & \delta^{(1)} + \delta^{(2)} = b(t)\left( \frac{d^2 \Phi_0}{dq^2}\right) + b^2(t)\left( \frac{d^2 \Phi_0}{dq^2}\right)^2 \nonumber \\
       & = & \frac{a_1^2+a_2^2}{2} + a_1 \bay{\lambda_1}{\alpha_1} +
       a_2 \bay{\lambda_2}{\alpha_2} \nonumber \\
       & + &  a_1^2 \bay{2 \lambda_1}{2
       \alpha_1}+ a_2^2 \bay{2
       \lambda_2}{2 \alpha_2} + a_1 a_2
       \bay{\lambda_1+\lambda_2}{\alpha_1+\alpha_2}+ a_1 a_2
       \bay{\lambda_1-\lambda_2}{ \alpha_1-\alpha_2}. \label{eq:2order}
\end{eqnarray}
Since the density contrast in 1D ZA can be expressed as a power series
of $d^2 \Phi_0/dq^2$ in Eq.~(\ref{eq:expansion}), the second-order term reflects the quadratic
density fields \cite{cb,wc}. This second-order term spawns modes with
wavenumbers that are from combinations of any 2 wavenumbers from the
parent ones, i.e. Fourier modes with wavenumbers $2\lambda_1$, $2\lambda_2$, $\lambda_1+\lambda_2$ and
$\lambda_1-\lambda_2$ spawned from combination of $\lambda_1$ and
$\lambda_2$. One important feature of quadratic density fields is that the
phases of the spawned modes follow the same kind of harmonic
relationship as the spawned wavenumbers, which we call hereafter
``wavenumber-phase harmonic relation''. Such relation subsequently forms phase
associations between Fourier modes and is crucial for bispectrum analysis. 

We can generalize the velocity potential in 1D ZA as a sum of cosine functions
$\Phi_0= - \sum_i A_i \cos( \lambda_i q + \alpha_i)$. The density
contrast of the first order and that up to the second order now become, respectively, 
\begin{equation}
\delta^{(1)} = \sum_i a_i \cos(\lambda_i+ \alpha_i) \equiv \sum_i a_i \bay{\lambda_i}{\alpha_i},
\end{equation}
where $a_i= b(t) A_i \lambda_i^2$, and
\begin{equation} 
\delta^{(1)} + \delta^{(2)}= \sum_i a_i \bay{\lambda_i}{\alpha_i} + \sum_{jk}
\frac{a_j a_k}{2} \left[\bay{\lambda_j - \lambda_k}{\alpha_j -
\alpha_k} + \bay{\lambda_j + \lambda_k}{\alpha_j + \alpha_k}
\right]. \label{eq:quadraticfield}
\end{equation}
If primordial Gaussianity is assumed, the phases
$\alpha_i$ of the initial velocity potential are uniformly random and
independently distributed between 0 and $2\pi$. Below we
categorize the effects induced by quadratic density fields.

\begin{itemize}
\item {\bf Mode spawning :} In Eq.(\ref{eq:quadraticfield}) beyond the
first order, new modes are spawned from the $\delta^{(2)}$ and
correlated phases are created following the spawned
modes. The wavenumbers of the spawned modes are formed from
combination of any 2 wavenumbers of the first order (the same as the
toy model) and the phases follow the {\it wavenumber-phase harmonic relation}. Some terms may appear
even at the earliest stage (when $b(t) A_j A_k (\lambda_j
\lambda_k/\lambda_i)^2 > 2 A_i$). Those are terms usually involving
high-frequency modes, i.e. high $\lambda_j$ or $\lambda_k$. Such phase
coupling can manifest itself through phase mapping \cite{mapping}. For
example, a short sequence of modes is formed with  a constant
difference in wavenumber $\Delta k=\lambda_k$: 
\begin{equation}
\frac{a_j a_k}{2}\bay{\lambda_j-\lambda_k}{\alpha_j - \alpha_k},
a_j \bay{\lambda_j}{\alpha_j}, \frac{a_j a_k}{2}
\bay{\lambda_j+\lambda_k}{\alpha_j + \alpha_k}, \label{eq:modespawning}
\end{equation}
where the middle mode is taken from the $\delta^{(1)}$. Due to the
wavenumber-phase harmonic relation, this phase sequence has a constant
phase difference $\Delta \phi =\alpha_k$ and can be mapped along a line parallel to
the diagonal through phase mapping technique. Such coupling between
modes with large $\Delta{\bbi k}$ is discussed in Chiang, Coles \&
Naselsky \cite{mapping}.  

\item {\bf Mode merging :} The newly--spawned modes from the
second order are not all independent but could merge with modes
of the same wavenumbers in the first order. Take the following two
modes from Eq.(\ref{eq:quadraticfield}) as an example, one from the
$\delta^{(1)}$ and the other from the $\delta^{(2)}$: 
\begin{equation}
 a_i\bay{\lambda_i}{\alpha_i}, \frac{a_j
 a_k}{2}\bay{\lambda_j-\lambda_k}{\alpha_j-\alpha_k}. \label{eq:modecoupling}
\end{equation} 
If $\lambda_j-\lambda_k =\lambda_i$, i.e. the wavenumber of the
newly-spawned mode coincides with the one in the first order, these two modes merge to form a new mode: 
\begin{equation}
\sqrt{a_i^2+\frac{a_j^2 a_k^2}{4}+a_i a_j a_k \cos(\alpha_j-\alpha_k
-\alpha_i)}\bay{\lambda_i}{\alpha_m},
\end{equation}
where
\begin{equation}
\alpha_m = \arctan \frac{a_i \sin \alpha_j + a_j a_k \sin(\alpha_j-\alpha_k) /2}{a_i \cos \alpha_j + a_j a_k \cos(\alpha_j-\alpha_k) /2}
\end{equation}
where the amplitude is modulated and the phase is
shifted. Such modulation in amplitudes and phases proceeds
gradually and the consequence is that the wavenumber-phase harmonic
relation is broken. Thus not all spawned
modes will enjoy the harmonic relation. Note
that in this case the evolution can still be at very early stage $b(t)
\ll 1$ as long as $\lambda_j \lambda_k/\lambda_i \gg (A_i/A_jA_k)^{1/2}$.  
\end{itemize}

\section{Full treatment of the 1D ZA}
To account for the interaction of the parent modes leading to mode
merging, we can give a full treatment of 1D ZA by starting from Eq.~(\ref{eq:delta1d}). We have 
\begin{eqnarray}
\delta_k & = & \frac{1}{2\pi}\int^{\pi}_{-\pi}
\left[\left(\frac{dx}{dq}\right)^{-1} - 1\right]e^{-\imath k x} dx  \nonumber \\
& = & \frac{1}{2\pi}\int^{\pi}_{-\pi}e^{-\imath kx} dq =
\frac{1}{2\pi}\int^{\pi}_{-\pi}\exp \{- \imath k [q+b(t)
u(q) ] \} dq, \label{eq:integral}
\end{eqnarray}
where $u(q)=-d\Phi_0(q)/d q$. We can give the same treatment to
Eq.~(\ref{eq:integral}) as Eq.~(\ref{eq:expansion}) by expressing it in
terms of a power series of $b(t)$: 
\begin{equation}
\delta_k=\sum^{\infty}_{p=0} \frac{(-\imath k b)^p}{p \; !}\frac{1}{2\pi}
\int^{\pi}_{-\pi} dq \; e^{-\imath kq} u(q)^p.
\end{equation}
Following Soda \& Suto \cite{sodasuto}, we can denote the Fourier
transform of $u(q)$ as $u_k$
\begin{equation}
\uk=\frac{1}{2\pi}\int^{\pi}_{-\pi} dq\; u(q) e^{-\imath k q},
\end{equation}
so the Fourier modes $\delta_k$ can be expressed as 
\begin{eqnarray}
\delta_k & = & -\imath kb \uk + \frac{(-\imath kb)^2}{2!}\sum_{k_1,k_2} u_{k_1} u_{k_2}
\delta^D[k-(k_1+k_2)] \nonumber \\
& + &  \frac{(-\imath kb)^3}{3!}\sum_{k_1,k_2,k_3} u_{k_1}
u_{k_2}u_{k_3} \delta^D[k-(k_1+k_2+k_3)]+\cdots, \label{eq:series}
\end{eqnarray}
where $\delta^D$ denotes Dirac-delta function. For 
$u(q)=-d\Phi_0/dq=- \sum_j A_j \lambda_j \sin(\lambda_j q +\alpha_j)$
from the velocity potential $\Phi_0=- \sum_j A_j \cos( \lambda_j q +\alpha_j)$,  
\begin{equation}
u_k = -\frac{1}{2\imath}\sum_j A_j \lambda_j e^{\imath \alpha_j}
\delta^D(\lambda_j -k)=-\frac{A_k \lambda_k}{2\imath} e^{\imath
\alpha_k}=\frac{A_k \lambda_k}{2} e^{\imath(\pi/2 + \alpha_k)}, \label{eq:uk}
\end{equation}
and $u_{-k}$ is simply its complex conjugate. 

Equation~(\ref{eq:series}) and (\ref{eq:uk}) provide us the insight into
mode spawning from the initial potential field $\Phi_0(q)$ and mode merging in
gravitational clustering. The Dirac-delta functions dictate the
spawning of Fourier modes, both the amplitudes and phases. If, for any
Fourier mode $k$, there is only one non-zero term in Eq.(\ref{eq:series}), either from the parent modes in the $\Phi_0(q)$ (i.e. from the 1st-order term $-\imath kb \uk$) or any combination from parent modes such that there is only one non-zero term in Eq.(\ref{eq:series}) triggered by the non-zero $\delta^D$, its phase $\alpha_k$ shall not
change and the amplitude grow steadily with $b(t)^{n}$ (according to
the order of this combination). If, on the other hand, there are more than one non-zero term in Eq.(\ref{eq:series}), indicating there is more than one set of parent-mode combination for the wavenumber $k$ [e.g. Eq.~(\ref{eq:modecoupling})], they will interact so that the phase
$\alpha_k$ will change accordingly. 

Take a toy model as an example, where the initial field has only two modes: 7 and 11. For the 1st-order term of Eq.(\ref{eq:series}), only $u_7$ and $u_{11}$ exist as the parent modes. For the 2nd-order term, those which are non-zero for $\delta^D$ are $k=7+7$, $7+11$, $11-7$, $11+11$; for the 3rd-order term, those which are non-zero are $k=7+7+7$, $11+11+11$, $7+7-11$, $7+7+11$, $11+11-7$ and $11+11+7$ \ldots etc.. The corresponding phases of these spawned modes come from multiplication of $u_7$ and $u_{11}$ and do not interact with each other.

However, if we have another toy model where the initial field has two modes, 7 and 14, again the parent modes are $u_7$ and $u_{14}$. Now for the Fourier mode $\delta_{k=14}$, for example, we have not only $b(t)\; u_{14}$ from the 1st-order term, but also $ b(t)^2\, u_{7}^2\, \delta^D[k-(7+7)]$ from the 2nd-order term, or even higher terms such as the 4th-order $b(t)^4 \, u_{7}^2\, u_{14}^2\, \delta^D[k-(14+14-7-7)]$ and the 5th-order $b(t)^5 \, u_{7}^4\, u_{14}\, \delta^D[k-(7+7+7+7-14)]$\ldots etc.. The evolution for $\delta_{k=14}$ is therefore from the vector addition of all these terms. Of course the higher order terms come into force and dominate at later times. One can then imagine that for an initial density distribution which has continuous Fourier modes, the modes spawned shall immediately interact with some parent modes, thereby complicating the phase and amplitude evolution in Fourier domain.

\section{Conclusion}
In this talk, I have used 1D ZA to illustrate some non-linear
effects induced by gravitational clustering: mode spawning
and mode coupling. The onset of
non-linearity can be illustrated by expanding the density contrast in terms of power series
of $b(t)$. The second-order of such expansion reflects quadratic
density fields, from which modes are spawned with phases following a
special wavenumber-phase harmonic relation. The wavenumber-phase harmonic relation holds as long as the wavenumbers of parent modes do not interact. In a full treatment of 1D ZA, such mode spawning and coupling take place not only during the departure from the linear regime, but also the entire gravitational clustering process. The complexity of Fourier amplitude and phase evolution in gravitational clustering comes from continuous interactions between mode merging and coupling with the spawned modes.

\end{document}